\documentclass[final,numberedheadings]{aipproc}   \layoutstyle{6x9}

\newcommand{\ignore}[1]{\relax}

\begin{document}

\title{Suppression of Shot-Noise in Quantum Cavities:
Chaos vs. Disorder
}


\author{Ph.~Jacquod}{
  address={D\'epartement de Physique Th\'eorique,
Universit\'e de Gen\`eve, CH-1211 Gen\`eve 4, Switzerland}
}
\author{Robert S.~Whitney}{
  address={D\'epartement de Physique Th\'eorique,
Universit\'e de Gen\`eve, CH-1211 Gen\`eve 4, Switzerland}
}

\begin{abstract}
We investigate the behavior of the shot-noise power through
quantum mechanical cavities in
the semiclassical limit of small electronic wavelength. In the absence of
impurity scattering, the Fano factor $F$,
giving the noise to current ratio, was previously found to disappear as more
and more classical, hence deterministic and noiseless 
transmission channels open
up. We investigate the behavior of $F$ as diffractive impurities are
added inside the cavity. We find that $F$ recovers its universal value
provided (i) impurities cover the full cavity so that
only a set of zero measure of classical trajectories may avoid them, and
(ii) the impurity scattering rate exceeds the inverse dwell time through the
cavity. If condition (i) is not satisfied, $F$ saturates 
below its universal value, even in the limit of strong scattering.
Our results corroborate the validity of the two-phase fluid
model according to which the electronic flow splits into two well separated 
components, a classical deterministic fluid and a stochastic
quantum-mechanical fluid. Only the latter carries shot-noise.
\end{abstract}

\maketitle


Time-resolved transport measurements through quantum mechanical
systems invariably observe current fluctuations, 
even in the (experimentally unrealistic) situation of a noiseless
measurement apparatus and at zero temperature. This intrinsically quantal
noise is usually referred to as shot-noise. It results from
the quantization of charge together with the statistical nature
of quantum mechanical transport \cite{Blan00}. 
As but one of the consequences of the quantum-classical correspondence
at large quantum numbers, it has been predicted
that, shot-noise through a chaotic ballistic cavity
disappears as the system becomes more 
and more classical, i.e. when the ratio of the electronic Fermi
wavelength to the linear cavity size vanishes 
$\lambda_{\rm F}/L \rightarrow 0$ \cite{Ben91}.
The purpose of this article is to discuss when and how shot-noise starts to be
reduced by an emergent classical, deterministic behavior.

Recent technological advances have made it 
possible to make electronic systems small and clean enough
that the resulting electronic mean free path is larger than 
the size of the confining potential defining the device \cite{Revdot1}.
The electronic motion in these quantum dots is thus ballistic, and
provided that their wavelength is short enough, the electrons have a
dynamics strongly related to the dynamics that a classical
particle would have. When this classical dynamics is chaotic, that is, when 
the shape of the dot differs significantly from a circle or an ellipse,
the transport properties are usually universal  
and well-captured by the Random Matrix Theory (RMT) 
of transport \cite{Ben97}. The starting point of RMT is the 
scattering
approach \cite{scatg}, which relates transport properties to the
system's scattering matrix 
\begin{eqnarray}\label{blocks}
{\cal S}= \left( \begin{array}{ll}
{\bf r} & {\bf t}' \\
{\bf t} & {\bf r}' 
\end{array}\right).
\end{eqnarray}
Here we consider a symmetric
two terminal geometry (the cavity is connected to two external leads
with equal number $N$ of propagating channels) for which  
${\cal S}$ is a 2-block by 2-block matrix, written in terms of 
$N \times N$ transmission 
(${\bf t}$ and ${\bf t}'$)
and reflection (${\bf r}$ and ${\bf r}'$) matrices. 
From ${\cal S}$, the system's conductance is 
given by $g={\rm Tr} ({\bf t}^\dagger {\bf t}) = \sum_n T_n$
($g$ is expressed in units of $e^2/h$ and the $T_n$'s are the
$N$ eigenvalues of ${\bf T}={\bf t}^\dagger {\bf t}$). 
RMT provides a statistical
theory of transport where ${\cal S}$ is assumed to be
uniformly distributed over one of Dyson's circular ensemble of
random matrices \cite{Meh91}.
Transport properties can be calculated from this sole assumption. 
For instance, within RMT, and in the limit $N \gg 1$
the transmission eigenvalues have a probability distribution 
\cite{Ben97}
\begin{equation}\label{probt}
P_{\rm RMT}(T) = \frac{1}{\pi} \frac{1}{\sqrt{T(1-T)}}
\end{equation}
for any $T \in[0,1]$. Note that classical particles would
be either deterministically transmitted, $T=1$ or reflected $T=0$.

The distribution of transmission eigenvalues is all one needs to get
fluctuations and higher moments of the current at low frequency.
The zero-frequency 
shot-noise power in particular is given by $S=2 e V \sum_n T_n (1-T_n)$
\cite{Blan00}. According to (\ref{probt}), $S$ 
is suppressed below its Poissonian value of 
$S_{\rm p}=2 e \langle I \rangle$ ($V$ is the applied voltage 
and $\langle I \rangle$ the time-averaged current)
by the Fano factor which reads
\begin{equation}\label{fanormt}
F = \frac{\sum_n T_n (1-T_n)}{\sum_n T_n} \;\;\;\;\;\;\;\; ; \;\;\;\;\;\;\;\;
 F_{\rm RMT}  = \frac{1}{4}.
\end{equation}
The RMT predictions (\ref{probt}) and (\ref{fanormt})
have been confirmed in various
transport experiments and numerical simulations
on open chaotic cavities \cite{Blan00,Ben97}.

\begin{figure}
\includegraphics[width=9cm]{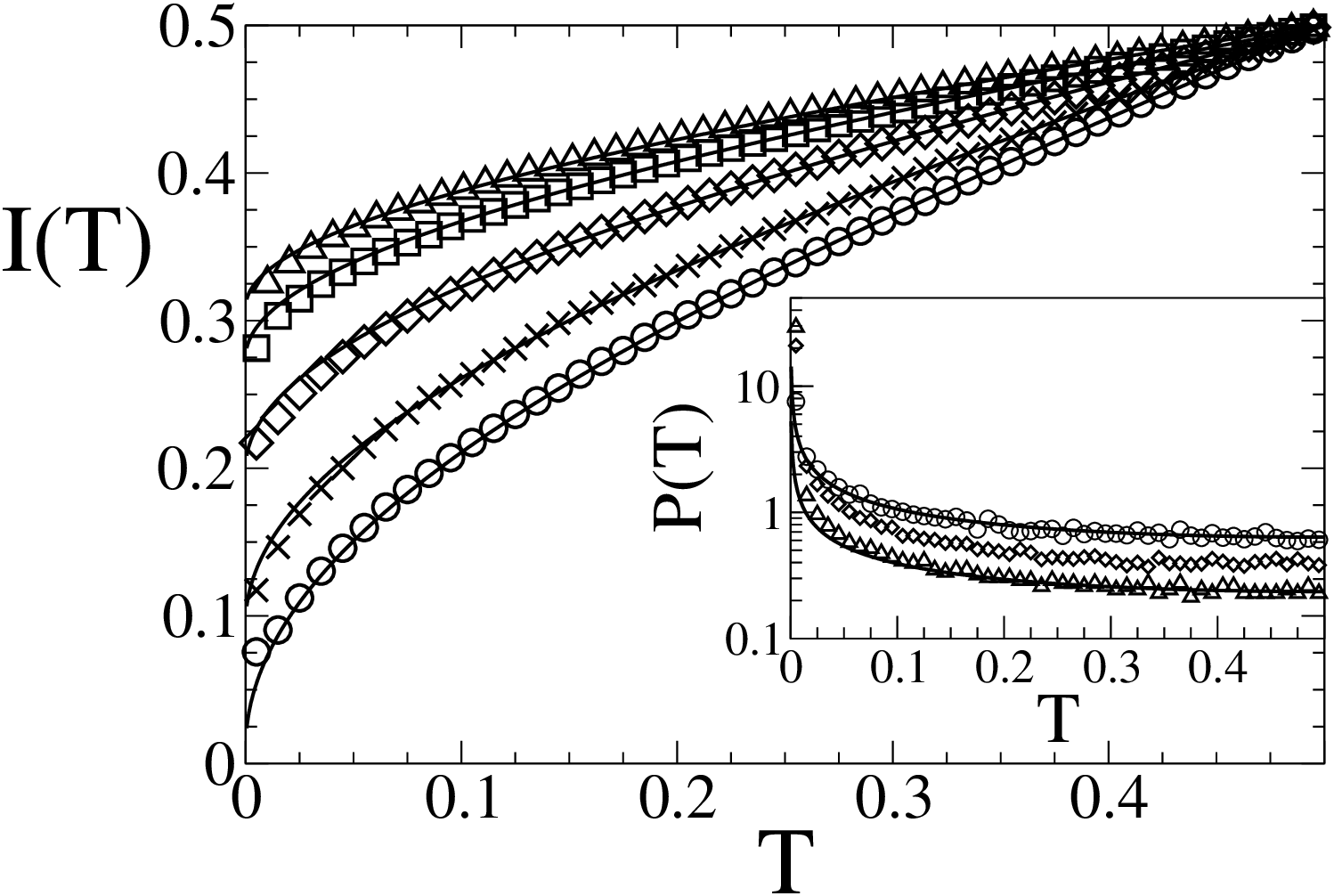}
\caption{\label{trans_prob1}
Integrated probability distribution $I(T)$
of transmission eigenvalues for 
$\tau_D=25$, $\hbar_{\rm eff}^{-1}=2048$ and $\tau_E \simeq 0$ (circles;
distribution calculated over 729 different samples); 
$\tau_D=5$, $\hbar_{\rm eff}^{-1}=128$ and 
$\tau_E=0.16$ ($\times$; 1681 samples), $\hbar_{\rm eff}^{-1}=1024$ and
$\tau_E=1.5$ (diamonds;  729 samples), $\hbar_{\rm eff}^{-1}=8192$ 
$\tau_E=2.8$ (squares; 16 samples), $\hbar_{\rm eff}^{-1}=65536$ 
and $\tau_E=4.1$
(triangles; 2 samples). The solid curves give the distribution $I_{\alpha}$ 
of Eq.~(\ref{iprobtalpha}), with $\alpha \approx 0.98$, 0.81,
0.6, 0.45, and 0.385 (from bottom to top).
Inset: Probability distribution $P(T)$ of transmission
eigenvalues for the same set of parameters as in the main panel
(squares and $\times$ have been removed for better visibility).
The solid curves 
give the universal distribution $P_{\rm RMT}$ of Eq.~(\ref{probt}) and
that of Eq.~(\ref{probtalpha}), with $\alpha = 0.39$ (from top
to bottom).
Note that $P(T)$ is symmetric around $T=0.5$.}
\end{figure}

In closed chaotic systems, the semiclassical limit 
$\lambda_{\rm F}/L \equiv \hbar_{\rm eff} \rightarrow 0$ 
usually results in a better and
better agreement with the Hamiltonian RMT of spectral fluctuations
\cite{Haake}. One may thus expect that the same applies
to transport in open systems. That is not so, as illustrated in
Fig.~\ref{trans_prob1}. The numerical data presented there 
(see also Ref.~\cite{Jac04}) show that 
instead of the RMT prediction of Eq.~(\ref{probt}),
the transmission eigenvalues appear
to be distributed according to 
\begin{equation}\label{probtalpha}
P_{\rm \alpha}(T) = \alpha P_{\rm RMT}(T) +\frac{1-\alpha}{2}
\left[\delta(T)+\delta(1-T)\right],
\end{equation}
with an increasingly deterministic behavior $\alpha \rightarrow 0$
as $\hbar_{\rm eff} \rightarrow 0$ (all classical parameters being fixed). 
The presence of $\delta$-peaks at $T=0$ and $T=1$ in 
$P_{\rm \alpha}(T)$ becomes evident once the integrated distribution
$I(T)=\int_0^T P(T') dT'$ is plotted. One has
\begin{equation}\label{iprobtalpha}
I_{\rm \alpha}(T) = \frac{2 \alpha}{\pi} \sin^{-1} \sqrt{T} \; + \;
\frac{1-\alpha}{2} (1 + \delta_{1,T}),
\end{equation}
so that $I_{\rm \alpha}(0)=(1-\alpha)/2$ vanishes only for
$\alpha=1$. It turns out that the parameter $\alpha$ is well approximated by
$\alpha \approx \exp(-\tau_e/\tau_d)$, in term of
the new time scale 
$\tau_e = -\lambda^{-1} \ln[\hbar_{\rm eff} \tau_d^2]$ and the
average dwell time $\tau_d$ through the cavity
\cite{Jac04}. In short, for a classically fixed configuration (i.e.
considering an ensemble of systems with fixed $\lambda$ and $\tau_d$),
the fraction
$(1-\alpha)/2$ of deterministic transmission eigenvalues $T=0,1$
increases as one goes deeper and deeper into the
semiclassical limit, $\hbar_{\rm eff} \rightarrow 0$. The rate of the 
crossover is set by a partially quantum, partially classical time scale, 
the {\it Ehrenfest time} $\tau_e$ \cite{Zas81}.
Compared to closed systems, the emergence of a finite $\tau_e$ 
has more profound an impact on transport properties once it
becomes comparable to the dwell time $\tau_d \propto
(N \hbar_{\rm eff})^{-1}$.

Inserting (\ref{probtalpha}) into (\ref{fanormt}) with the numerically
extracted value $\alpha \approx \exp(-\tau_e/\tau_d)$,
one directly recovers $F \propto
\exp[-\tau_e/\tau_d]$, in agreement with the analytical prediction of
Refs.~\cite{Agam00,Sil03}, the 
experimental results of Ref.~\cite{Ober02} and the numerical data 
of Ref.~\cite{Two03}. They can be qualitatively understood 
by first realizing that complex quantum systems split into the two
classes of quantum chaotic and quantum disordered systems \cite{Ale96}.
Ballistic cavities in particular belong
to the first class, for which electronic wavepackets are carried along 
very few classical paths until
the time $\tau_e$ after which they have a finite
probability to be found on trajectories that their center of mass
would not follow classically. This dynamical
diffraction process restores quantum mechanical stochasticity
for larger times, however it does not affect short trajectories
with $\tau<\tau_e$. Thus, in quantum chaotic systems, two
classes of classical trajectories emerge, depending on their
dwell time through the cavity. Short trajectories with $\tau<\tau_e$
are able to carry an electronic wavepacket deterministically through the
cavity (i.e. with transmission probability $T=0$ or 1). If the electronic
wavepacket sits on longer trajectories with $\tau > \tau_e$ on the
other hand, diffraction splits it into pieces before its exit,
and quantum mechanical stochasticity (with $T \in ]0,1[$) prevails.
The fraction of scattering
trajectories in the stochastic
subset is obtained via the dwell time distribution $\rho(\tau)$ by
\begin{equation}
\alpha \equiv \int_{\tau_e}^\infty \rho(\tau) {\rm d}\tau.
\end{equation} 
In a chaotic system one has
$\rho(\tau)=\tau_d^{-1} \exp[-\tau/\tau_d]$, hence the
fraction of stochastically transmitted channels gives
$\alpha \approx  \exp[-\tau_e/\tau_d]$ for the
weight $\alpha$ of Eq.~(\ref{probtalpha}), in agreement with the
numerics shown in Fig.~\ref{trans_prob1}.
This is the essence of the {\it two-phase-fluid model} \cite{Jac04},
originally proposed in Ref.~\cite{Sil03} and given a microscopic 
foundation in Ref.~\cite{wj2004}

Fig.~\ref{trans_prob1} provides us with a direct evidence for
the validity of the two-phase fluid. Other evidences of this kind
have been found in investigations of the excitation spectrum of 
Andreev billiards, i.e. ballistic cavities in contact with a 
superconductor \cite{Goo05}.
In a previous work \cite{wj2004}, 
we used an approach based on a semiclassical expansion
for the Green's function in term of a sum over classical
trajectories. Together with the construction of a quantum-mechanical
phase-space basis, this 
allowed us to import classical concepts such as Liouville conservation and
determinism into quantum mechanics, thereby providing with a microscopic
foundation for the two-phase fluid model \cite{comment}. Our purpose
in the reminder of this article is to check numerically the model by
investigating its quantum chaos -- quantum disorder crossover.

\begin{figure}
\includegraphics[width=9cm]{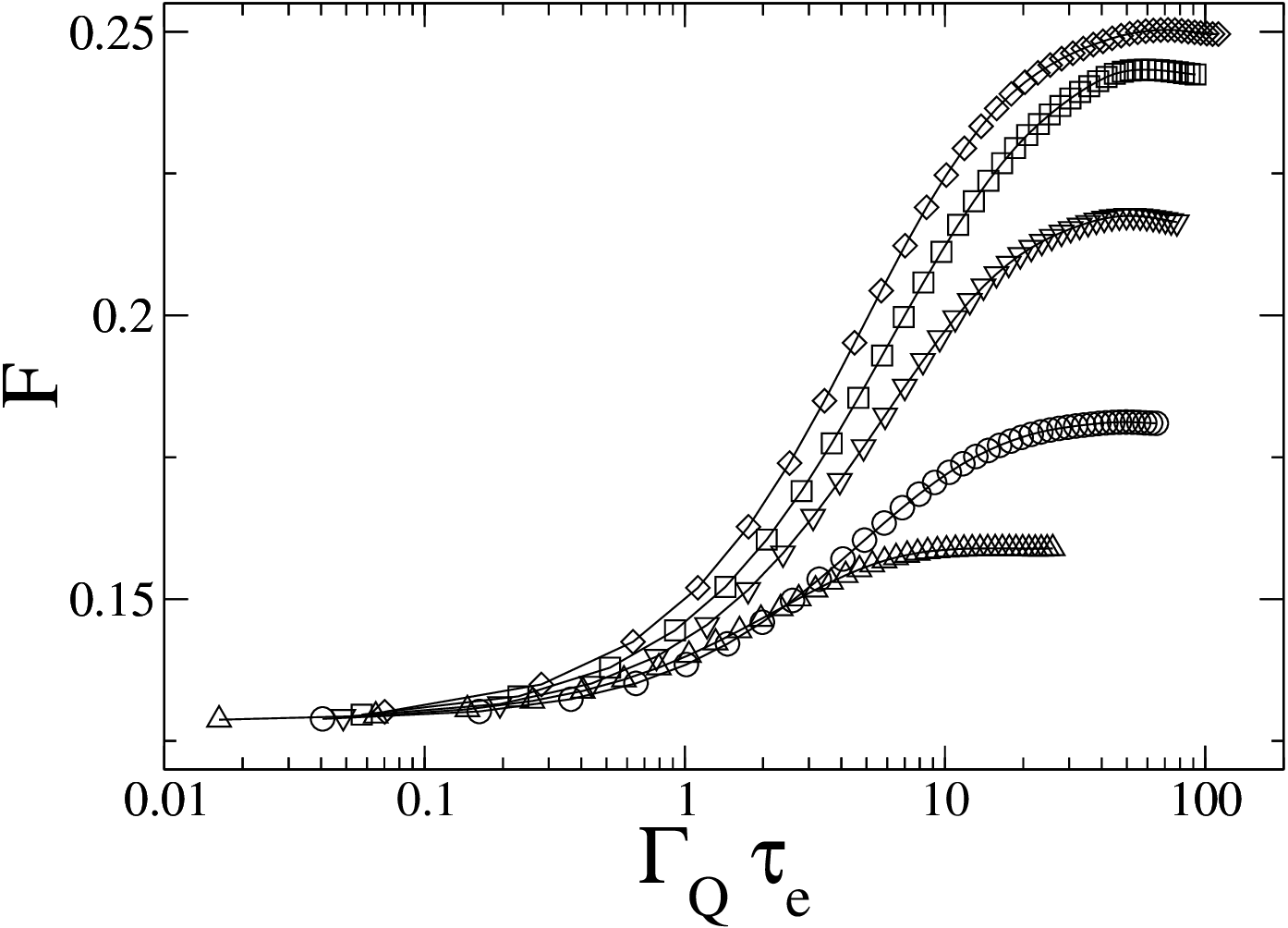}
\caption{\label{xoverF}
Crossover from quantum chaotic behavior (left) to quantum disordered behavior
(right) of the Fano factor $F$ as a function of $\Gamma_Q \tau_e$.
Each set of data corresponds to an average over four different samples
with $\hbar_{\rm eff}^{-1}=8192$, $K=7.65$,
$\tau_D=5$ and $\xi=1250$ (upward triangles), 
2500 (circles), 3750 (downward triangles),
5000 (squares) and 7500 (diamonds).}
\end{figure}

To this end, we use the open kicked rotator model and follow the procedure
described e.g. in Ref.~\cite{Two03}. So far, the model has
been implemented only in its quantum chaotic version. Here we add a diffraction
term to it so that its Floquet (time-evolution) operator has matrix elements
\begin{eqnarray}\label{kickedU}
U_{m,m'} &=& \hbar_{\rm eff}^{1/2} \;
e^{-(iK/4\pi\hbar_{\rm eff})[\cos(2\pi m \hbar_{\rm eff})+\cos(2\pi m'
\hbar_{\rm eff})]}  \;
e^{-(i/4\pi\hbar_{\rm eff}) [\eta(m)+\eta(m')]} 
\nonumber \\
&& \times \sum_l e^{2 \pi i l(m-m')\hbar_{\rm eff}} 
e^{-(\pi i \hbar_{\rm eff}/2) l^{2}},
\end{eqnarray}
with a randomly distributed function $ \langle \eta(m) \eta(m') \rangle =
u^2 \; \delta_{m,m'} \; \Theta(|m-m_0|-\xi/2)$. Such a point-like
diffractive impurity potential corresponds to the extreme quantum
disorder limit. The length scale $\xi$ allows to cover all or only a part
of the system, and $u$ sets the strength of the impurity scattering. 
We extracted the scattering
rate $\Gamma_Q$ from the width of the Local Spectral Density of States
(LDOS) induced by $\eta(m)$ for the closed kicked rotator. As is commonly the
case \cite{Jac95}, we found that the LDOS has a Lorentzian shape with
a width $\Gamma_Q \approx 0.016 \; u^2 \sqrt{\xi}/\hbar_{\rm eff}$
in a wide range of parameter.
One expects that, as $\Gamma_Q$ becomes comparable to the Ehrenfest time,
quantum diffraction effects start to eat away the determinism of 
short trajectories.

\begin{figure}
\includegraphics[width=9cm]{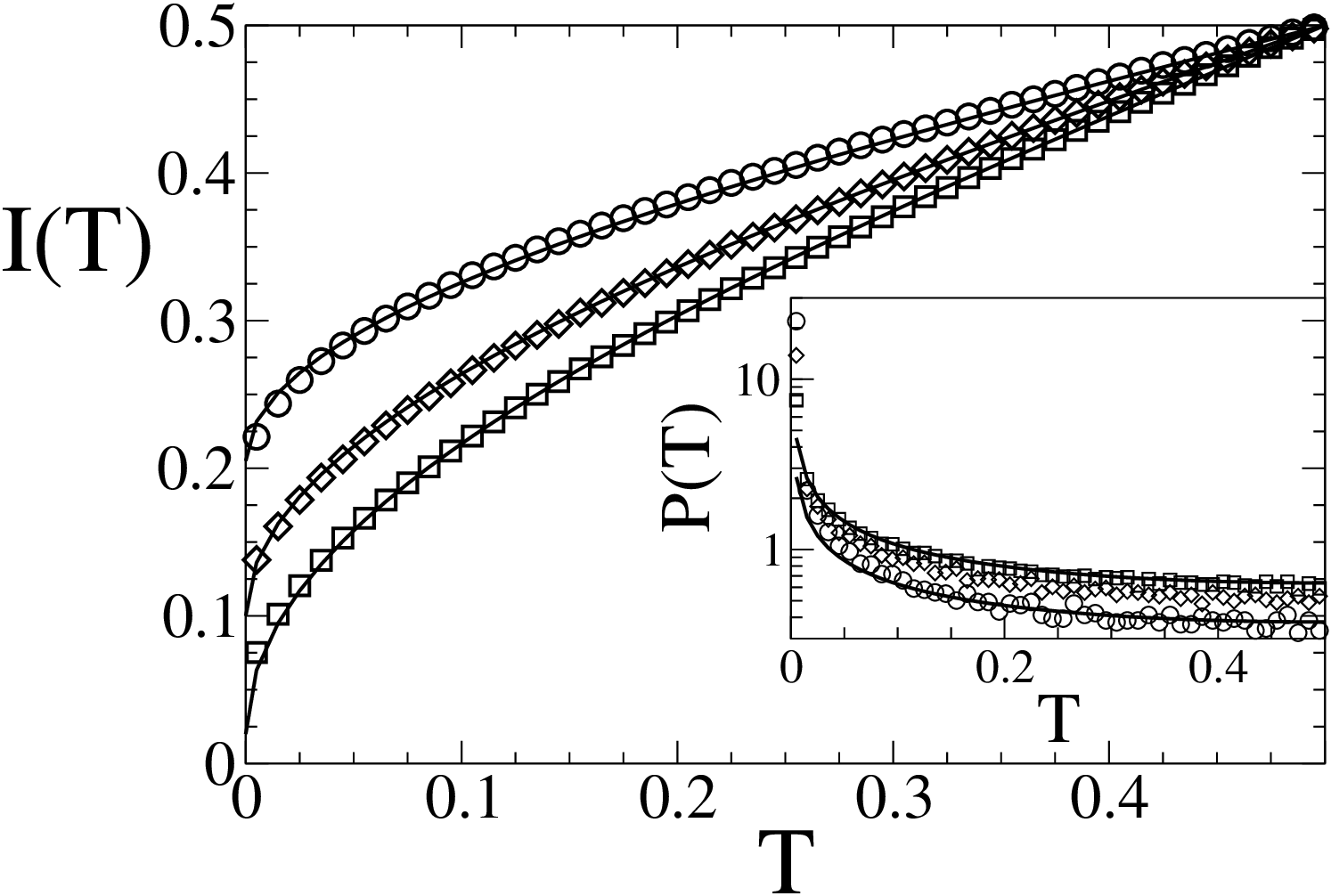}
\caption{\label{trans_prob2}
Integrated probability distribution $I(T)$
of transmission eigenvalues for the open kicked rotator
model with diffractive disorder of Eq.~(\ref{kickedU}). 
The data have been obtained from 16 different samples with
parameters $\hbar_{\rm eff}^{-1}=8192$, $K=7.65$
$\tau_D=5$ and $u=0.4$, with $\xi=0$ (circles), 2500 (diamonds), 
and 7500 (squares). The solid curves give the distribution $I_{\alpha}$ 
of Eq.~(\ref{iprobtalpha}), with $\alpha \approx 0.99$, 0.76, and 0.52
from bottom to top.
Inset: Probability distribution $P(T)$ of transmission
eigenvalues for the same set of parameters as in the main panel.
The solid curves 
give the universal distribution $P_{\rm RMT}$ of Eq.~(\ref{probt}) and
that of Eq.~(\ref{probtalpha}), with $\alpha = 0.59$ respectively.}
\end{figure}

We first show in Fig.~\ref{xoverF} the behavior of the Fano factor as
the diffractive scattering rate is cranked up, all other
parameter being fixed. For $\Gamma_Q=0$, $F<0.25$ lies below its universal
value, indicating a finite $\tau_e$, thus a finite fraction of
non-diffractive scattering orbits. This fraction gets reduced once
$\Gamma_Q$ increases, and accordingly $\partial F/\partial \Gamma_Q > 0$.
The most striking feature of Fig.~\ref{xoverF}, however, is that 
as long as the diffractive disorder does not cover the whole 
system volume, i.e. for $\xi < 
\hbar_{\rm eff}^{-1} (1-\tau_d^{-1})$, $F$ saturates below its
universal value, even in the limit $\Gamma_Q \tau_e \rightarrow \infty$.
This reflects the fact that some short trajectories are able to avoid the
diffractive potential and thus remain deterministic. The existence of
two separated phase-space fluids allows only that part of the deterministic
fluid which directly scatters off the impurities to become stochastic.
It is not clear from our numerics if,
for $\xi >\hbar_{\rm eff}^{-1} (1-\tau_d^{-1})$,
one has an exponential or a power-law behavior of the Fano
factor as predicted in
Ref.~\cite{Ober02,Bul05}. More detailed 
investigations are necessary to draw definite conclusions. 

We finally show on Fig.~\ref{trans_prob2} the behavior of the
distribution of transmission eigenvalues in the saturated regime
$\Gamma_Q \tau_e \gg 1$. Clearly, the distribution and integrated
distribution follow Eqs.~(\ref{probtalpha}) and (\ref{iprobtalpha}).
In contrast to the quantum chaotic case, 
we qualitatively found a dependence
$\alpha \propto \exp[-\tau_e/\tau_d] \xi \hbar_{\rm eff}/
(1-\tau_d^{-1})$, again reflecting a reduction of that part of
the deterministic component which directly touches 
the diffractive potential. 

All these findings support the two-phase fluid hypothesis, that is, the
splitting of the cavity into a stochastic and a deterministic cavity.
The latter being noiseless, shot-noise is suppressed by a factor
reflecting its phase-space measure relative to the total phase-space.
This measure is reduced by the presence of diffractive disorder, however, 
only homogeneously spread impurities are able to diffract all trajectories,
thus only in this case does one recover universality at large diffractive
scattering rate. This complement recent investigations of shot-noise 
with homogeneously spread diffractive disorder in regular
cavities \cite{Aig05}.

This work has been supported by the Swiss National
Science Foundation.

\end{document}